**Title:**
Timing of ancient human Y lineage depends on the mutation rate: A comment on Mendez et al.

**Author:**
Melissa A. Wilson Sayres*[1]

**Affiliation:**
1. Departments of Statistics and Integrative Biology, University of California, Berkeley, Berkeley, California 94720 United States of America

**Correspondence:**
*mwilsonsayres@berkeley.edu

**Letter**
Mendez et al.[1] recently report the identification of a Y chromosome lineage from an African American that is an outgroup to all other known Y haplotypes, and report a time to most recent common ancestor, TMRCA, for human Y lineages that is substantially longer than any previous estimate [2-4]. The identification of a novel Y haplotype is always exciting, and this haplotype, in particular, is unique in its basal position on the Y haplotype tree. However, at 338 (237-581) thousand years ago, kya, the extremely ancient TMRCA reported by Mendez et al.[1] is inconsistent with the known human fossil record (which estimate the age of anatomically modern humans at 195 +- 5 kya [5]), with estimates from mtDNA (176.6 +- 11.3 kya [6], and 204.9 (116.8-295.7) kya [7]) and with population genetic theory. The inflated TMRCA can quite easily be attributed to the extremely low Y chromosome mutation rate used by the authors [1].

The mutation rate is not identical across chromosome types. Male mutation bias refers to the higher rate of mutation in the male lineage versus the female lineage, and is thought to result primarily from the higher number of rounds of replication in the male germline relative to the female germline [8; 9]. Because the Y chromosome is present only in the male germline, its mutation rate is expected to reflect the male mutation rate, while the mutation rate on the autosomes, present half of the time in the male germline, and half of the time in the female germline, should reflect the sex-average mutation rate [9]. The magnitude of male mutation bias increases with increasing generation time [10]. In humans, specifically, the relevance of male mutation bias is manifest in the increasing mutation rate observed with increasing paternal age [11].

Mendez et al.[1], assuming there is a direct adjustment for the male-specific mutation bias, derive an estimate of $6.17 \times 10^{-10}$ mutations/nucleotide/year (range: $4.39 \times 10^{-10} - 7.07 \times 10^{-10}$) on the Y chromosome, assuming a median of 30 years per generation, using only information about autosomal substitution rates [11]. But, recent work, comparing estimates of male mutation bias using different chromosome comparisons (X/Autosome, Y/Autosome, X/Y), have shown that while male mutation bias may explain most of the differences in the mutation rates between each sex chromosome and the autosomes, it cannot account for all of the rate variation [10; 12]. Thus, estimates of the mutation rate on the Y chromosome based on autosomal rates may not be accurate, even when accounting for variance in the paternal age at conception.

The mutation rate Mendez et al.[1] calculate for the Y chromosome is very low, and is actually quite similar to estimates of the mutation rate on the autosomes from recent resequencing projects ($0.4 \times 10^{-9} - 0.6 \times 10^{-9}$ mutations/nucleotide/year; summarized nicely by [13]). Mendez et al.[1] claim that they are being cautious by obtaining estimate of the mutation rate on the Y by using rates from human pedigrees, because it is likely more accurate to estimate the mutation rate using pedigree information than comparative genomics. However, the authors do not use the mutation rate for the human Y chromosome, computed from a deep-rooted human pedigree, as reported in 2009 [14]. Xue et al.[14] report a mutation rate on the human Y chromosome of $1.0 \times 10^{-9}$ mutations/nucleotide/year (95% CI: $3.0 \times 10^{-10} - 2.5 \times 10^{-9}$). As Mendez et al.[1] report themselves, using the higher mutation rate ($1.0 \times 10^{-9}$) results in a more reasonable estimate of the TMRCA of 209 kya [1]. This "reduced" estimate still highlights the novelty of the newly identified Y chromosome to increase TMRCA estimates for the Y compared to what other large-scale studies of Y chromosomes have identified (e.g., 101-115 kya [3]; and 142 kya [4]).

Further, under neutral population genetic theory, a TMRCA of 338 kya for the Y is certainly not in line with TMRCA estimates for the mtDNA [6; 7], and is also inconsistent with TMRCA estimates for the X [15; 16], and for autosomes [17]. Under neutral expectations, the effective population size of the Y chromosome is expected to be equal to the effective population size of the mtDNA, one quarter that of the autosomes, and one third that of the X chromosome. Current observations of the TMRCA across these other genomic regions are not compatible with the high Y chromosome TMRCA that Mendez et al. [1] report using their derivation of the Y chromosome mutation rate, but are consistent with a Y chromosome TMRCA calculated using the mutation rate estimated from a Y-pedigree [14] (Table 1). In addition, our recent work has shown that the observed diversity on the entire Y chromosome is approximately one tenth what is expected, due to the effects of selection acting to reduce diversity on this non-recombining chromosome (http://arxiv.org/abs/1303.5012). If selection is acting to reduce diversity on the Y, then the TMRCA estimates of Mendez et al. [1] are likely substantial under-estimates, putting them even more at odds with estimates of the TMRCA on the mtDNA, X, and autosomes.

Mendez et al. [1] postulate that the extremely old TMRCA they calculate could indicate archaic introgression, or may be suggestive of a highly structured ancestral human population. Both of these things may be true in the history of anatomically modern humans, but a much simpler explanation for their observations is that the mutation rate used to estimate the TMRCA for the Y chromosome was simply too low.

**Acknowledgements**
I would like to thank Kirk Lohmueller and Rasmus Nielsen for comments on the manuscript. MAWS is supported by the Miller Institute for Basic Research in Science.

**Table 1. Expected and observed TMRCA for autosomes, X chromosome, and mtDNA, under different Y chromosome TMRCAs.** For a given TMRCA on the Y chromosome, the expected TMRCA for the autosomes, X chromosome and mtDNA can be computed, assuming that the effective population size of the Y is equal to that of the mtDNA, one third that of the X chromosome, and one quarter that of the autosomes. The expected TMRCAs are calculated using the TMRCA for the Y ($TMRCA_Y$) computed using the low Y chromosome mutation rate derived from an autosomal mutation rate ($6.17 \times 10^{-10}$ mutations/nucleotide/year; $TMRCA_Y$ = 338 kya), and using the mutation rate estimated from a Y-chromosome pedigree ($1.0 \times 10^{-9}$ mutations/nucleotide/year; $TMRCA_Y$ = 209 kya).

|  | Expected TMRCA (kya) | | Observed TMRCA (kya) | Citation |
|---|---|---|---|---|
|  | $TMRCA_Y$ = 338 | $TMRCA_Y$ = 209 | | |
| Autosomes | 1352 | 836 | 796 | 17 |
| X chromosome | 1014 | 627 | 741 (± 168) | 16 |
|  |  |  | 535 (±119) | 15 |
| mtDNA | 338 | 209 | 204.9 (116.8-295.7) | 7 |
|  |  |  | 176.6 (±11.3) | 6 |